\documentclass[12pt]{article}
\usepackage{amssymb}

\def\theorem#1{\begin{center}
\begin{minipage}{.95\textwidth}{\bf Theorem\ } \em #1 
\end{minipage}\end{center} } 
\def\corollary#1{\begin{center}
\begin{minipage}{.95\textwidth}{\bf Corollary\ } \em #1 
\end{minipage}\end{center} }  

\begin{document}

\begin{center}
{\Large\bf Triality, Biquaternion and Vector\\[0.4cm]
Representation of the Dirac Equation}\\[0.8cm]
Liu Yu-Fen \\[0.2cm]
{\it Institute of Theoretical Physics}, {\it Academia Sinica,}\\[0.1cm]
{\it \ Beijing 100080, China}\\[0.1cm]
email: liuyf@itp.ac.cn\\[0.3cm]
(September, 2001)
\end{center}

\vspace*{0.4cm}

\begin{abstract}
The triality properties of Dirac spinors are studied, including a
construction of the algebra of (complexified) biquaternion. It is proved
that there exists a vector-representation of Dirac spinors. The massive
Dirac equation in the vector-representation is actually self-dual. The
Dirac's idea of non-integrable phases is used to study the behavior of
massive term.
\end{abstract}

\section{Introduction}

The first example of a vector representation for spinors was derived by E.
Cartan$^{[1]}$. He noticed that the group $Spin(2n)$ is the double covering
group of the rotation group $SO(2n)$, i.e. this group has two basic
half-spinor (semi-spinor) representations of degree $2^{n-1}$. In the
special case {\it only} when $2n=2^{n-1}$, the group $Spin(8)$ has just
three irreducible representations of degree $8$, all real, and the three
representation spaces (vector space) $R$, (semi-spinors spaces) $S_{+}$ and
$S_{-}$ are, remarkably, on an equal footing. It turns out that there is an
extra automorphism, known as `triality'$^{[1][2][3]}$, which changes the
spinor representations of $SO(8)$ to the vector representation and is not
related to any symmetry of other $SO(2n)$ groups. The word 'triality' is
applied to the algebraic and geometric aspects of the $\Sigma _3$ symmetry
which $Spin(8)$ has. One wonders what three objects does the symmetric group
$\Sigma _3$ permutes; and the answer is that it permutes representations.

\begin{center}
\begin{minipage}{.95\textwidth}
{\bf Theorem }\  \em (Cartan's principle of triality$^{[1][2]}$): There
exists an automorphism $J$
of order 3 of the vector space $A=R\times S_{+}\times S_{-}$ 
(dimension=8+8+8) which has following properties: $J$ leaves the quadratic
form $\Omega $ and the cubic form $F$ invariant; $J$ maps $R$ onto $S_{+}$, %
$S_{+}$ onto $S_{-}$, and $S_{-}$ onto $R$.
\end{minipage}
\end{center}

The law of composition in the algebra $A$, is defined in terms of the
quadratic forms $\Omega $ and cubic form $F$ only. For the $SO(8)$ case the
defined algebra is the algebra of Cayley octonions$^{[2]}$, and it is clear
that any automorphism of the vector space $A$ which leaves the quadratic
forms $\Omega $ and cubic form $F$ invariant is an automorphism of this
algebra. The Brioschi's formula in the real domain (i.e. The product of
two sums of eight squares is itself the sum of eight squares.), which
deduced from the above
considerations by Cartan$^{[1]}$, can be considered as a normed condition
for division octonions. Generally, any division algebra gives a triality$%
^{[4][5]}$, and it follows that normed trialities only occur in dimensions
1,2,4 or 8. (Here we use a generalized concept of 'triality' advocated by
Adams$^{[4]}$.) This conclusion is quite deep. By comparison, Hurwitz's
classification of normed division algebras (in the real domain) is easier
to prove$^{[6]}$.

The construction of the algebras from trialities has tantalizing links to
physics. In the Standard Model, all particles other than the Higgs boson
transform either as vectors or spinors. The interaction between matter
(quarks and leptons) and the forces (gauge bosons) is described by a {\it %
trilinear} map involving two spinors and a vector. It is fascinating that the
same sort of mathematics can be used both to construct a suitable algebra
and to describe the interaction between matter and forces.

One prima facie problem with the above speculation is that physics uses
spinors associated to Lorentz groups rather than $SO(8)$ rotational groups,
due to the fact that 4-dimension space-time has a Lorentzian rather than
Euclidean metric. Luckily the dimension of spinor space, depends on both
the dimension of vector space and
the {\it signature} of the metric. In 4-Lorentzian dimensions, the gamma
matrices $\gamma ^\mu $ are (at least) $4\times 4$, and thus the number of 
{\it complex} spinor components are four too, or equivalently 4+4
independent real components. Thus, we have here three spaces each of four
dimensions, that of vectors, that of semi-spinors of the first type, and
that of semi-spinors of the second type. The quadratic form in the vector
space is defined by means of $\eta _{\mu \nu }=diag(+1,-1,-1,-1)$, and the
quadratic form in the spinor space is defined by means of Dirac conjugate
spinors ($\overline{\psi }\psi $ here $\overline{\psi }=\psi ^{*T}\gamma ^0$
). We will prove that there exists a double covering vector representation
of Dirac spinors, and that there exists an automorphism $J$ of order 3 in
the $A=M^{1+3}\times S_{+}^4\times S_{-}^4$ (vector space and two
half-spinor spaces), which leaves quadratic forms and the special cubic form
invariant up to the sign. (In Minkowski space we are faced with spacelike
and timelike vectors, thus sometimes it is convenient to use the term {\it %
pseudoscalar}, here the prefix '{\it pseudo}' referring to automorphism $J$%
.) The situation differs from the case of $SO(8)$, and we propose to
denominate this symmetry as a ``{\it ding}'' construction. (Chinese {\it %
ding} is an ancient vessel which has two loop handles and three legs. It is
the metaphor of tripartite balance of forces.) The algebra associated with
the above construction is the algebra of (complexified) {\it bi}quaternion.
In this sense the vector representation of Dirac spinor is equivalent to a
biquaternion representation.

Moreover we will prove that the massive Dirac equation in the vector
representation is actually self-dual.

It is important to stress that the triality transformation, i.e. the
representations permutation, is not a symmetry of the theory. It maps one
description of the theory to {\it another} description of the same theory.

\section{Triality and ''Ding'' construction}

Let us first introduce trinomial unit-basis $(\varphi ^\beta ,f^\alpha
,j^\mu )$, where the basic unit vector $j^\mu $ and two basic unit spinors $%
\varphi ^\beta $, $f^\alpha $ are normalized so that
\begin{equation}
\begin{array}{lllll}
\varphi =j_\mu i\gamma ^\mu f & , & f=j_\mu i\gamma ^\mu \varphi & , & j^\mu
=-\overline{\varphi }i\gamma ^\mu f=\overline{f}i\gamma ^\mu \varphi
\end{array}
\end{equation}
\begin{equation}
\begin{array}{c}
\overline{\varphi }\varphi =-\overline{f}f=-j^\mu \eta _{\mu \nu }j^\nu =1 \\ 
\overline{\varphi }f=\overline{f}\varphi =j_\mu (\overline{\varphi }\gamma
^\mu \varphi )=j_\mu (\overline{f}\gamma ^\mu f)=0
\end{array}
\end{equation}
Here any two objects determine the third. (Raising and lowering 
indices by Lorentzian metric $\eta ^{\mu \nu }$ and $\eta _{\mu \nu }$, the
summation convention is assumed for repeated indices.)

By using the quantities $(j_\mu i\gamma^\mu)$, $(i\gamma^\mu f)$,
$(i\gamma^\mu \varphi)$ we can translate between different types of vectors
and spinors indexes. In addition, we can
introduce another unit vector $k^\mu $, which is determined by the above
trinomial unit-basis, that will play a very important role in our theory. 
\begin{equation}
\begin{array}{lllll}
k^\mu \stackrel{def.}{=}\overline{\varphi }\gamma ^\mu \varphi =\overline{f}%
\gamma ^\mu f & , & k^\mu k_\mu =1 & , & k^\mu j_\mu =0
\end{array}
\end{equation}
\begin{equation}
\begin{array}{lll}
k_\mu \gamma ^\mu f=-f & , & k_\mu \gamma ^\mu \varphi =\varphi
\end{array}
\end{equation}
The existence of such objects in Dirac theory is verified by the special
case (28) below.

Furthermore we need the following algebraic properties
\begin{equation}
\begin{array}{lllll}
\overline{\varphi }\gamma ^\mu \gamma ^\nu \varphi & = & -\overline{f}\gamma
^\mu \gamma ^\nu f & = & \eta ^{\mu \nu }+i\epsilon ^{\mu \nu \lambda \rho
}k_\lambda j_\rho \\ 
\overline{\varphi }\gamma ^\mu \gamma ^\nu f & = & \overline{f}\gamma ^\mu
\gamma ^\nu \varphi & = & i(k^\mu j^\nu -j^\mu k^\nu )
\end{array}
\end{equation}
\begin{equation}
\begin{array}{ccc}
\overline{\varphi }\gamma ^\mu \gamma ^\nu \gamma ^\lambda \varphi & =%
\overline{f}\gamma ^\mu \gamma ^\nu \gamma ^\lambda f & =i\epsilon ^{\mu \nu
\lambda \rho }j_\rho +t^{\mu \nu \lambda \rho }k_\rho \\ 
-\overline{\varphi }\gamma ^\lambda \gamma ^\nu \gamma ^\mu f & =\overline{f}%
\gamma ^\mu \gamma ^\nu \gamma ^\lambda \varphi & =\epsilon ^{\mu \nu
\lambda \rho }k_\rho -it^{\mu \nu \lambda \rho }j_\rho
\end{array}
\end{equation}
here $\epsilon ^{\mu \nu \lambda \rho }$ is the Levi-Civita symbol, with the
definition $\epsilon ^{0123}=1$, and 
\begin{equation}
\begin{array}{l}
\epsilon ^{\mu \nu \lambda \rho }=\frac i4tr(\gamma ^5\gamma ^\mu \gamma
^\nu \gamma ^\lambda \gamma ^\rho ) \\ 
t^{\mu \nu \lambda \rho }\stackrel{def.}{=}\frac 14tr(\gamma ^\mu \gamma
^\nu \gamma ^\lambda \gamma ^\rho )=(\eta ^{\mu \nu }\eta ^{\lambda \rho
}+\eta ^{\mu \rho }\eta ^{\nu \lambda }-\eta ^{\mu \lambda }\eta ^{\nu \rho
})
\end{array}
\end{equation}
(Comments: In calculating S-matrix elements and transition rates for
processes involving particles of spin $\frac 12$, we often encounters
traces of products of Dirac gamma matrices. The above symbols
$\epsilon ^{\mu \nu \lambda \rho }$ and $t^{\mu \nu \lambda \rho }$ are not
new for physics, we always use them in all such calculations.)

We are now in a position to construct the suitable algebra, that are
needed. Let
\begin{equation}
\begin{array}{lll}
c^{\mu \nu \lambda } & \stackrel{def.}{=}(t^{\mu \nu \lambda \rho
}-i\epsilon ^{\mu \nu \lambda \rho })k_\rho & =(c^{\lambda \nu \mu })^{*} \\ 
\check{c}^{\mu \nu \lambda } & \stackrel{def.}{=}(t^{\mu \nu \lambda \rho
}-i\epsilon ^{\mu \nu \lambda \rho })j_\rho & =(\check{c}^{\lambda \nu \mu
})^{*} \\ 
c_5^{\mu \lambda } & \stackrel{def.}{=}-c^{\nu \lambda \mu }j_\nu =\check{c}%
^{\nu \lambda \mu }k_\nu & =-c_5^{\lambda \mu }
\end{array}
\end{equation}
(in which asterisk $*$ represents complex conjugate.) then 
\begin{equation}
\begin{array}{l}
c^{\mu \nu \sigma }\eta _{\sigma \delta }(c^{\delta \rho \lambda
})^{*}+c^{\mu \rho \sigma }\eta _{\sigma \delta }(c^{\delta \nu \lambda
})^{*}=2\eta ^{\mu \lambda }\eta ^{\nu \rho } \\ 
\check{c}^{\mu \nu \sigma }\eta _{\sigma \delta }(\check{c}^{\delta \rho
\lambda })^{*}+\check{c}^{\mu \rho \sigma }\eta _{\sigma \delta }(\check{c}%
^{\delta \nu \lambda })^{*}=-2\eta ^{\mu \lambda }\eta ^{\nu \rho }
\end{array}
\end{equation}
\begin{equation}
\begin{array}{c}
c_5^{\mu \rho }\eta _{\rho \sigma }c_5^{\sigma \nu }=\eta ^{\mu \nu } \\ 
\check{c}^{\mu \nu \lambda }=-c^{\mu \nu \sigma }\eta _{\sigma \rho
}c_5^{\rho \lambda }=(c_5^{\mu \sigma })^{*}\eta _{\sigma \rho }c^{\rho \nu
\lambda }
\end{array}
\end{equation}
The Dirac operators are defined as 
\begin{equation}
\begin{array}{lll}
\check{D}^{\mu \sigma }\stackrel{def.}{=}\check{c}^{\mu \nu \sigma }\partial
_\nu & , & (\check{D}^{\mu \lambda })^{*}\eta _{\lambda \rho }(\check{D}%
^{\rho \nu })=-\eta ^{\mu \nu }\square \\
D^{\mu \sigma }\stackrel{def.}{=}c^{\mu \nu \sigma }\partial
_\nu & , & (D^{\mu \lambda })^{*}\eta _{\lambda \rho }(D%
^{\rho \nu })=\eta ^{\mu \nu }\square 
\end{array}
\end{equation}
We define a bilinear law of composition $\otimes $ (and $\check{%
\otimes}$) for complex vectors 
\begin{equation}
\begin{array}{ccc}
(G\otimes H)^\lambda \stackrel{def.}{=}G_\mu c^{\mu \lambda \nu }H_\nu
=S^\lambda & , & (G\check{\otimes}H)^\rho \stackrel{def.}{=}G_\mu \check{c}%
^{\mu \rho \nu }H_\nu
\end{array}
\end{equation}
The dot product is defined by means of Minkowski metric, i.e. $G\cdot H%
\stackrel{def.}{=}G^\mu \eta _{\mu \nu }H^\nu $, and is not necessarily
real; let alone positive.

We can prove that 
\begin{equation}
\begin{array}{l}
(G\otimes H)\otimes K=G\otimes (H\otimes K) \\ 
(G\check{\otimes}H)\check{\otimes}K=G\check{\otimes}(H\check{\otimes}K)
\end{array}
\end{equation}
and 
\begin{equation}
\begin{array}{l}
(G\cdot G)(H\cdot H)=(G\otimes H)\cdot (G\otimes H) \\ 
(G\cdot G)(H\cdot H)=-(G\check{\otimes}H)\cdot (G\check{\otimes}H)
\end{array}
\end{equation}
The algebra defined here is associative and noncommutive. The last
identities look like the modified normed condition. (It is worth a passing
to mention that a somewhat vague formulation of normed condition for
{\it reals} is  
$\left(\sum\limits_{i=1}^n a_i^2\right)\left(\sum\limits_{i=1}^n
b_i^2\right)=\left(\sum\limits_{i=1}^n c_i^2\right)$ in which $c_i=%
\sum\limits_{i=1}^n \sum\limits_{k=1}^n a_j\gamma_{jik}b_k$ 
where $\gamma _{ijk}$ are constants. It is natural to
ask whether or not do analogous identities involving more squares exist.
It has occupied the mind of mathematicians for many years. Only in 1898 did
Hurwitz$^{[6]}$ prove that the identities of interest to us are possible
{\it only} for $n=1,2,4,8$.
This result, of course, is intimately related to the well known
result, due to ...... that there exist only four normed algebras over the 
{\it reals} : {\bf R}, {\bf C}, {\bf H}, {\bf O}.)

Because the algebra above is associative, it can be considered in terms of
the matrices. Let 
\begin{equation}
\begin{array}{ccc}
G=G_\mu e^\mu & ; & (e^\mu )_{\cdot \lambda }^{\nu \cdot }\stackrel{def.}{=}%
c^{\nu \sigma \mu }\eta _{\sigma \lambda }
\end{array}
\end{equation}
here $G_\mu $ are complex and the matrices $e^\mu $ can be considered as the
hypercomplex basic units. We can prove that a bilinear law of composition
(12) is equivalent to matrix multiplication 
$(G_\mu e^\mu )_{\cdot \sigma }^{\lambda \cdot}(H_\nu e^\nu )_{\cdot
\rho}^{\sigma \cdot} =(S_\mu e^\mu )_{\cdot \rho }^{\lambda \cdot }$.
In the special coordinate system (28),
where $k^\mu =\delta ^{\mu 0}$, the basis element $e^0=I$ and
for $\nu =1,2,3$, we have $e^\nu =\sqrt{-1}\hat{e}^\nu $ here 
\begin{equation}
\hat{e}^1=\left( 
\begin{array}{cccc}
0 & -i & 0 & 0 \\ 
-i & 0 & 0 & 0 \\ 
0 & 0 & 0 & -1 \\ 
0 & 0 & 1 & 0
\end{array}
\right)
\end{equation}
\begin{equation}
\hat{e}^2=\left( 
\begin{array}{cccc}
0 & 0 & -i & 0 \\ 
0 & 0 & 0 & 1 \\ 
-i & 0 & 0 & 0 \\ 
0 & -1 & 0 & 0
\end{array}
\right) _{}
\end{equation}
\begin{equation}
\hat{e}^3=\left( 
\begin{array}{cccc}
0 & 0 & 0 & -i \\ 
0 & 0 & -1 & 0 \\ 
0 & 1 & 0 & 0 \\ 
-i & 0 & 0 & 0
\end{array}
\right) _{}
\end{equation}
One notices that they obey the multiplication law : 
\begin{equation}
\begin{array}{c}
\hat{e}^1\hat{e}^1=\hat{e}^2\hat{e}^2=\hat{e}^3\hat{e}^3=\hat{e}^1\hat{e}^2%
\hat{e}^3=-I \\ 
\hat{e}^1\hat{e}^2=-\hat{e}^2\hat{e}^1=\hat{e}^3
\end{array}
\end{equation}
(the last relation being cycle). These relations define a structure of the
algebra. We shall call this algebra the algebra of {\it complexified}
{\it bi}quaternions$^{[7]}$.

Furthermore we can construct the commutative Jordan algebra in the following
way 
\begin{equation}
G\circ K\stackrel{def.}{=}G_\mu (t^{\mu \rho \nu \sigma }k_\sigma )K_\nu
=\frac 12(G\otimes K+K\otimes G)
\end{equation}
We can prove that 
\begin{equation}
\begin{array}{c}
G\circ K=K\circ G \\ 
\lbrack (G\circ G)\circ K]\circ G=(G\circ G)\circ (K\circ G)
\end{array}
\end{equation}

Now let us study the problem of the vector representation
of Dirac spinors. The most important thing to us is : the trinomial
unit-basis $(\varphi ^\beta ,f^\alpha ,j^\mu )$ which satisfies the above
conditions exists. And it is easily verified from the special case (28)
below.

We know that a Dirac spinor has 4 complex components, or equivalently 4+4=8
independent real components. Any Dirac spinor can be decomposed into the sum
of two `half-spinors' $\Psi =\Psi _1+\Psi _2$. Each half-spinor has 4
independent real components and can be constructed by means of $f$ and
$\varphi$, i.e. 
\begin{equation}
\begin{array}{lll}
\Psi _1=B^\mu \eta _{\mu \nu }i\gamma ^\nu f & , & \Psi _2=N^\mu \eta _{\mu
\nu }i\gamma ^\nu \varphi
\end{array}
\end{equation}
\begin{equation}
\begin{array}{lll}
B^\mu & =\frac 12(\overline{f}i\gamma ^\mu \Psi -\overline{\Psi }i\gamma
^\mu f) & =\frac 12(\overline{f}i\gamma ^\mu \Psi _1-\overline{\Psi }%
_1i\gamma ^\mu f) \\ 
N^\mu & =\frac 12(\overline{\Psi }i\gamma ^\mu \varphi -\overline{\varphi }%
i\gamma ^\mu \Psi ) & =\frac 12(\overline{\Psi }_2i\gamma ^\mu \varphi -%
\overline{\varphi }i\gamma ^\mu \Psi _2)
\end{array}
\end{equation}
Here $B^\mu $ and $N^\mu $ are real vectors. They define the vector
representation of the half-spinors $\Psi _1$ and $\Psi _2$.

Sometime it is convenient to pass from trinomial unit-basis $%
(\varphi ^\beta ,f^\alpha ,j^\mu )$ to the null-basis $(r^\alpha ,l^\beta
,k_{\pm }^\mu )$, where the normalized ''right-handed'' spinor $r$ and
''left-handed'' spinor $l$ (pure spinors) are determined in the following
way
\begin{equation}
\begin{array}{lllll}
\overline{r}l=\overline{l}r=2 & , & f=\frac i2(r-l) & , & \varphi =\frac
12(r+l) \\ 
k_{\pm }^\mu =\frac 12(k^\mu \pm j^\mu ) & , & k_{\pm }^\mu \eta _{\mu \nu
}k_{\pm }^\nu =0 & , & k_{\pm }^\mu \eta _{\mu \nu }k_{\mp }^\nu =\frac 12
\end{array}
\end{equation}
It is easy to prove that 
\begin{equation}
\begin{array}{ccc}
k_\mu \gamma ^\mu r=k_\mu ^{-}\gamma ^\mu r=l & , & k_\mu ^{+}\gamma ^\mu r=0
\\ 
k_\mu \gamma ^\mu l=k_\mu ^{+}\gamma ^\mu l=r & , & k_\mu ^{-}\gamma ^\mu l=0
\end{array}
\end{equation}
The Dirac spinor can be equivalently decomposed into the sum of a
''right-handed'' and a ''left-handed'' spinors $\Psi =R+L$, here 
\begin{equation}
\begin{array}{ccc}
R\stackrel{def.}{=}\frac 12G^\mu \eta _{\mu \nu }\gamma ^\nu l & , &
L\stackrel{def.}{=}\frac{-1}2G^{*\mu}\eta _{\mu \nu }\gamma ^\nu r
\end{array}
\end{equation}
\begin{equation}
G^\mu =\frac 12(\overline{r}\gamma ^\mu \Psi -\overline{\Psi }\gamma ^\mu
l)=B^\mu +iN^\mu
\end{equation}

In order to understand our idea easily, it is convenient to work in a
special coordinate system such that 
\begin{equation}
\begin{array}{ll}
\lbrack \varphi ^\beta ]^T & =[1,0,0,0] \\ 
\lbrack f^\alpha ]^T & =[0,0,i,0] \\ 
j^\mu & =(0,0,0,1) \\ 
k^\mu & =(1,0,0,0)
\end{array}
\end{equation}
(here we use the Dirac representation of gamma matrices). In this special
case 
\begin{equation}
\Psi =\Psi _1(B)+\Psi _2(N)=\left( 
\begin{array}{l}
B^3+iN^0 \\ 
B^1+iB^2 \\ 
B^0+iN^3 \\ 
-N^2+iN^1
\end{array}
\right) .  \label{26}
\end{equation}

Now let $V^\mu \in M^{1+3}$ be a vector in Minkowski space, $\Psi _1(B)\in
S_1$ and $\Psi _2(N)\in S_2$ are two half-spinors referred to the trinomial
unit-basis defined above. The quadratic forms and the special cubic-form are
defined as 
\begin{equation}
\begin{array}{lllll}
V^\mu \eta _{\mu \nu }V^\nu =V_\nu V^\nu & , & \overline{\Psi }_1\Psi
_1=-B_\nu B^\nu & , & \overline{\Psi }_2\Psi _2=N_\mu N^\mu
\end{array}
\end{equation}
\begin{equation}
V^\mu \eta _{\mu \nu }(\overline{\Psi }_1\gamma ^\mu \Psi _2+\overline{\Psi}%
_2\gamma ^\mu \Psi _1)=2(\epsilon _{\nu \lambda \rho \sigma }k^\sigma )V^\nu
N^\lambda B^\rho
\end{equation}
One realizes that there exists a ''ding'' automorphism $J$ of order 3 in $%
M^{1+3}\times S_1\times S_2$, which leaves quadratic two-forms (30) and the
trilinear-form (31) invariant up to the sign. $J$ maps $M$ onto $S_1$, $S_1$
onto $S_2$, and $S_2$ onto $M$. (Or equivalently $V\rightarrow B\rightarrow
N\rightarrow V$.)

\section{Bosonization of Dirac equation.}

The most interesting for physicists is: by passing from ordinary spinor
representation to the vector representation, one can expresses Dirac
Lagrangian in the Bosonic form 
\begin{equation}
\begin{array}{ll}
& \frac 12[\overline{\Psi }i\gamma ^\mu (\partial _\mu -ieA_\mu )\Psi
-((\partial _\mu +ieA_\mu )\overline{\Psi })i\gamma ^\mu \Psi ]-m\overline{%
\Psi }\Psi \\ 
= & \frac 12[(\nabla_\mu G_\nu)^{*} i\check{c}^{\nu \mu \lambda} G_\lambda
-G_\nu ^{*} i\check{c}^{\nu \mu \lambda } \nabla_\mu G_\lambda 
+m(G_\nu ^{*}G^{*\nu}+G_\nu G^\nu )]
\end{array}
\end{equation}
Here $\nabla_\mu G_\lambda \stackrel{def.}{=} \partial_\mu G_\lambda
-ieA_\mu \eta _{\lambda \rho} c_5^{\rho \sigma} G_{\sigma}{ }$ 
 and  $\check{c}^{\nu \mu \lambda }$ ,  $c_5^{\rho \lambda }$
are defined by (8).

It is important to notice that from the abstract point of view, there is an
arbitrariness in our Bosonization procedure since it depends on the choice
of the neutral elements $j^\mu $ end $k^\mu $. In the Lagrangian (32) they
are arbitrary but must be fixed. The only restrictions on them are that
$k_\mu k^\mu =-j_\mu j^\mu =$ $1$ and $k_\mu j^\mu =0$. The remaining basis
elements ($c^{\mu \nu \sigma }$, $\check{c}^{\nu \mu \rho }$, $c_5^{\mu \nu
} $, ......) can be reconstructed if desired, once $j^\mu $ and $%
k^\mu $ have been chosen.

The corresponding massive Dirac equation in the vector-representation takes
the form of 
\begin{equation}
\check{c}^{\mu \nu \lambda }i\nabla_\nu G_\lambda -mG^{\mu *}=0{.}
\end{equation}
Or equivalently in the following self-dual form 
\begin{equation}
\begin{array}{l}
\partial _\mu G^\mu -imj_\mu G^{\mu *}=0 \\ 
G_{\mu \nu }=\frac i2\ \epsilon _{\mu \nu \lambda \rho }G^{\lambda \rho }
\end{array}
\end{equation}
here (for simplicity) we take $A_\mu =0$ and 
\begin{equation}
G_{\mu \nu }\stackrel{def.}{=}[(\partial _\mu G_\nu +imj_\mu G_\nu
^{*})-(\partial _\nu G_\mu +imj_\nu G_\mu ^{*})]
\end{equation}

The Dirac equation (33) can be rewritten in the real form 
\begin{equation}
\begin{array}{l}
\check{\epsilon}^{\mu \nu \lambda }\partial _\nu B_\lambda -\check{t}^{\mu
\nu \lambda }\partial _\nu N_\lambda -eA_\nu (t^{\mu \nu \lambda }B_\lambda
+\epsilon ^{\mu \nu \lambda }N_\lambda )-mB^\mu =0 \\ 
\check{\epsilon}^{\mu \nu \lambda }\partial _\nu N_\lambda +\check{t}^{\mu
\nu \lambda }\partial _\nu B_\lambda -eA_\nu (t^{\mu \nu \lambda }N_\lambda
-\epsilon ^{\mu \nu \lambda }B_\lambda )+mN^\mu =0
\end{array}
\end{equation}
(here $\check{\epsilon}^{\mu \nu \lambda }=\check{\epsilon}^{\mu \nu \lambda
\rho }j_\rho $ and $\epsilon ^{\mu \nu \lambda }=\epsilon ^{\mu \nu \lambda
\rho }k_\rho $) or equivalently 
\begin{equation}
\begin{array}{l}
\partial _\mu N^\mu -[mB^\mu +eA_\nu (B_\lambda t^{\mu \nu \lambda
}+N_\lambda \epsilon ^{\mu \nu \lambda })]j_\mu =0 \\ 
\partial _\mu B^\mu -[mN^\mu -eA_\nu (N_\lambda t^{\mu \nu \lambda
}-B_\lambda \epsilon ^{\mu \nu \lambda })]j_\mu =0 \\ 
(\nabla_\mu^\prime N_\nu -\nabla_\nu^\prime N_\mu )=\frac 12\epsilon _{\mu
\nu \lambda \rho }(\nabla ^{\prime \lambda} B^\rho -\nabla^{\prime \rho}
B^\lambda )
\end{array}
\end{equation}
here 
\begin{equation}
\begin{array}{l}
\nabla_\mu^\prime N_\nu \stackrel{def.}{=}\partial_\mu N_\nu +\frac m2
\check{\epsilon}_{\mu \nu \rho }N^\rho +e(j_\mu \eta _{\nu \sigma }\epsilon
^{\sigma \lambda \rho }-\frac 12\check{\epsilon}_{\mu \nu \sigma }t^{\sigma
\lambda \rho })A_\lambda N_\rho \\ 
\nabla_\mu^\prime B_\nu \stackrel{def.}{=}\partial_\mu B_\nu -\frac m2
\check{\epsilon}_{\mu \nu \rho }B^\rho +e(j_\mu \eta _{\nu \sigma }\epsilon
^{\sigma \lambda \rho }-\frac 12\check{\epsilon}_{\mu \nu \sigma }t^{\sigma
\lambda \rho })A_\lambda B_\rho
\end{array}
\end{equation}

In addition, let us define the operator $\nabla_\mu^c \stackrel{def.}{=}%
(\partial _\mu +imj_\mu C^{*})$, such that 
\begin{equation}
\nabla _\mu^c G_{\nu \lambda }\stackrel{def.}{=}\partial_\mu G_{\nu \lambda
}+imj_\mu G_{\nu \lambda }^{*}
\end{equation}
where $C^{*}$ is the operator of complex conjugation: $C^{*}\Phi =\Phi ^{*}$%
. In this notation the identity 
\begin{equation}
\nabla_\mu^c G_{\nu \lambda }+\nabla_\lambda^c G_{\mu \nu }+\nabla_\nu^c
G_{\lambda \mu }=0
\end{equation}
looks like the Bianchi identity, and 
\begin{equation}
\begin{array}{ll}
& \frac 14[G_{\mu \nu }\epsilon ^{\mu \nu \lambda \rho }G_{\lambda \rho
}+G_{\mu \nu }^{*}\epsilon ^{\mu \nu \lambda \rho }G_{\lambda \rho }^{*}] \\ 
= & \frac 12\partial _\mu [\epsilon ^{\mu \nu \lambda \rho }(G_\nu
G_{\lambda \rho }+G_\nu ^{*}G_{\lambda \rho }^{*})] \\ 
= & 2\partial _\mu [\epsilon ^{\mu \nu \lambda \rho }(B_\nu \partial
_\lambda B_\rho -N_\nu \partial _\lambda N_\rho +2mB_\nu N_\lambda j_\rho )]
\end{array}
\end{equation}
looks like the Chern-Pontryagin density and a total derivative of the
Chern-Simons density. Thus the Dirac Lagrangian can be modified by the
additional total derivative of the Chern-Simons density introduced above.

\section{Lorentz rotations, U(1) and Chiral transformations}

Because of the lack of commutativity there are two types of $\check{\otimes}$
products in the biquaternion algebra, the left and right multiplications.
Thus for a
given biquaternion $G$ (and $q$ such that $q_\mu q^\mu =-1$) we can consider
the two maps, i.e. two transformations$^{[7]}$: $\widetilde{G}=q\check{%
\otimes}G$ and $\widetilde{G}=G\check{\otimes}q$, depending as we multiply $%
q $ on the left or the right. In the vector representation, these
transformations take the form of 
\begin{equation}
\begin{array}{l}
\widetilde{G^\mu }=G^\nu (\eta _{\nu \lambda }\check{c}^{\lambda \mu \sigma
}q_\sigma )=G^\nu [S_1 (q)]_{\nu \cdot }^{\cdot \mu } \\ 
\widetilde{H^\mu }=(q_\sigma \check{c}^{\sigma \mu \lambda }\eta _{\lambda
\nu })H^\nu =[S_2 (q)]_{\cdot \nu }^{\mu \cdot } G^\nu 
\end{array}
\end{equation}
From (modified normed condition) (14) we find that under the above
transformations the dot product $G_\mu \eta ^{\mu \nu }G_\nu $ is
unchanged. Thus the two-form of the spinors $\overline{\Psi }\Psi
=- \frac 12(G_\nu^{*}G^{*\nu }+G_\nu G^\nu )$ is invariant under the above
'complex rotations'. (Similarly, if
$q_\mu q^\mu =1$, then $\check{c}^{\nu \mu \lambda }$ is
replaced by $c^{\mu \nu \lambda }$).

For the space-time (real) vector $x^\mu $ (and $\partial _\mu $) the Lorentz
transformation, in quaternionic terms, is characterized by the above
unit-biquaternion $q$ through the relation (considered as left-right mixed
transformations$^{[7]}$): 
\begin{equation}
\widetilde{x^\mu }=[\Lambda (q)]_{\cdot \nu}^{\mu \cdot } x^\nu
=[S_2 (q^{*})]_{\cdot \sigma }^{\mu \cdot }
(x^\nu [S_1 (q)]_{\nu \cdot }^{\cdot \sigma })
\end{equation}
The constrain $q_\mu q^\mu =-1$ (implying two
real conditions) ensures that this transformation is 6-dimensional. This
transformation preserves the value of the dot product and the reality of the
space-time vector. Thus we have here two types of vectors with one-to-one
distinct transformation operations $[S_1(q)]_{\nu \cdot }^{\cdot \mu }$ and 
$[\Lambda (q)]_{\cdot \nu }^{\mu \cdot }$. In contrast with an ordinary real
space-time vector, we shall use the name 's-vector' for $G^\mu $ which
transforms as (42).

We can prove that under the above Lorentz transformations (with fixed
$j^\mu$ and $k^\nu $) 
\begin{equation}
\begin{array}{l}
\lbrack \Lambda (q)]_{\cdot \rho }^{\nu \cdot}
[S_2(q^{*})]_{\cdot \sigma}^{\mu \cdot}
\check{c}^{\sigma \rho \delta }[S_1(q)]_{\delta \cdot }^{\cdot \lambda}
=\check{c}^{\mu \nu \lambda } \\ 
\lbrack \Lambda (q)]_{\cdot \rho }^{\nu \cdot}
[S_2(q^{*})]_{\cdot \sigma}^{\mu \cdot}
c^{\sigma \rho \delta }[S_1(q)]_{\delta \cdot }^{\cdot \lambda}
=c^{\mu \nu \lambda }
\end{array}
\end{equation}
and this ensures the Lorentz invariance of the Dirac Lagrangian (32) which
is written in the s-vector form.

Also the Lagrangian (32) is invariant under $U(1)$ gauge transformations 
\begin{equation}
\begin{array}{lll}
\widetilde{\Psi }=\Psi e^{i\alpha }=\Psi (\cos \alpha +i\sin \alpha ) & , & 
\widetilde{A_\mu }=A_\mu +\partial _\mu \alpha
\end{array}
\end{equation}
In the vector representation it is equivalent to 
\begin{equation}
\begin{array}{l}
\widetilde{B_\mu }=B_\mu \cos \alpha -[\check{\epsilon}_{\mu \nu \lambda
}k^\lambda B^\nu -(k_\mu j_\nu -k_\nu j_\mu )N^\nu ]\sin \alpha \\ 
\widetilde{N_\mu }=N_\mu \cos \alpha -[\check{\epsilon}_{\mu \nu \lambda
}k^\lambda N^\nu +(k_\mu j_\nu -k_\nu j_\mu )B^\nu ]\sin \alpha
\end{array}
\end{equation}
or in complex form 
\begin{equation}
\widetilde{G_{}^\mu }=(\eta ^{\mu \nu }\cos \alpha +ic_5^{\mu \nu }\sin
\alpha )G_\nu =(\exp i\alpha c_5)^{\mu \nu }G_\nu ^{}
\end{equation}
It looks like a chiral transformation for $G^\mu $, here $(c_5^{\mu \sigma
}\eta _{\sigma \rho }c_5^{\rho \nu })=\eta ^{\mu \nu }$, and from which the
De Moivre theorem is deduced $(\cos \alpha +ic_5\sin \alpha )^n=(\cos
n\alpha +ic_5\sin n\alpha )$.

We know that the massless Dirac Lagrangian is invariant under the chiral
transformation 
\begin{equation}
\begin{array}{lll}
\Psi \rightarrow e^{ia\gamma _5}\Psi & ; & \overline{\Psi }\rightarrow 
\overline{\Psi }e^{ia\gamma _5}
\end{array}
\end{equation}
In the s-vector representation it is equivalent to 
\begin{equation}
G_\mu \rightarrow e^{ia}G_\mu
\end{equation}
It means that a chiral transformation for $\Psi $ is equivalent to a $U(1)$
transformation for $G_\mu $. The mass term $ m\overline{\Psi }\Psi
=-\frac m2 (G_\nu ^{*}G^{*\nu }+G_\nu G^\nu )$ is not invariant under the 
above chiral transformation, but is invariant under $U(1)$ transformation
(45)-(47). Therefore we must distinguish between the plan wave solutions
for $\Psi $ and for $G^\mu$.

Now, we define the ''dual'' transformation 
\begin{equation}
\begin{array}{lllll}
B\rightarrow N & , & N\rightarrow -B & , & m\rightarrow -m
\end{array}
\end{equation}
which leaves the quadratic form $m\overline{\Psi }\Psi =m(N_\mu N^\mu -B_\nu
B^\nu )$ , the cubic form $V_\mu B_\nu N_\lambda \epsilon ^{\nu \mu \lambda
} $ and thus the above Lagrangian and Dirac equation invariant. (It looks
like the electro-magnetic duality which is accompanied by $e\dashrightarrow
q $ and $q\dashrightarrow -e$). In this sense the half-spinor of the first
type is dual to the half-spinor of the second type.

One may have many choices of "representations" of the neutral elements
$(f,\varphi ,j^\mu ,k^\mu )$ and $\check{c}^{\mu \nu \lambda}$. For
example, we can change the representations in the following way
\begin{equation}
\begin{array}{lll}
\widetilde{f}=\frac 12(\overline{a}+\frac 1a)f+\frac i2(\overline{a}-\frac
1a)\varphi  & , & \widetilde{r}=\overline{a}r \\
\widetilde{\varphi }=\frac 12(\overline{a}+\frac 1a)\varphi -\frac i2(%
\overline{a}-\frac 1a)f & , & \widetilde{l}=\frac 1al
\end{array}
\end{equation}
\begin{equation}
\begin{array}{l}
\widetilde{j^\mu }=\frac 12(a\overline{a}+\frac 1{a\overline{a}})j^\mu
+\frac 12(a\overline{a}-\frac 1{a\overline{a}})k^\mu  \\
\widetilde{k^\mu }=\frac 12(a\overline{a}+\frac 1{a\overline{a}})k^\mu
+\frac 12(a\overline{a}-\frac 1{a\overline{a}})j^\mu
\end{array}
\end{equation}
here $\overline{a}$ represents the complex conjugate of $a$. Correspondingly
\begin{equation}
\widetilde{G^\mu }=[\frac 12({a}+\frac 1a)\eta ^{\mu \nu }+\frac
12({a}-\frac 1a)c_5^{\mu \nu }]G_\nu
\end{equation}
The Dirac Lagrangian (32) is invariant under this change of representation.
In fact the $U(1)$ transformation (47), can be considered as the special
case of (53).

\section{The behavior of the mass term.}
 
\theorem{There exists a suitable complex vector $K^\mu $ associated with 
spinor $\Psi
=R+L$ (and completely determined by it) such that 
\begin{equation}
\begin{array}{ccc}
K_\mu \gamma ^\mu R=K_\mu ^{-}\gamma ^\mu R=L & , & \overline{R}K_\mu
^{*}\gamma ^\mu =\overline{L} \\ 
K_\mu \gamma ^\mu L=K_\mu ^{+}\gamma ^\mu L=R & , & \overline{L}K_\mu
^{*}\gamma ^\mu =\overline{R}
\end{array}
\end{equation}
The explicit form of these vectors are 
\begin{equation}
\begin{array}{l}
K_\mu =\frac{\overline{R}\gamma _\mu R}{2\overline{R}L}+\frac{\overline{L}%
\gamma _\mu L}{2\overline{L}R}=K_\mu ^{+}+K_\mu ^{-} \\ 
K_\mu \eta ^{\mu \nu }K_\nu =1
\end{array}
\end{equation}
}
(Comments: the existence of such vector is due to the existence of $k_\mu
=k_\mu ^{+}+k_\mu ^{-}$ which has been introduced in the previous section
(see eq.(25)), and thus has close relation with the ''ding'' construction.)

\corollary{The mass quadratic form equal to the interaction trilinear form 
\begin{equation}
\begin{array}{c}
\overline{R}L=K_\mu \overline{R}\gamma ^\mu R=K_\mu ^{*}\overline{L}\gamma
^\mu L \\ 
\overline{L}R=K_\mu \overline{L}\gamma ^\mu L=K_\mu ^{*}\overline{R}\gamma
^\mu R
\end{array}
\end{equation}
If $\Psi =R+L$ , then 
\begin{equation}
m\overline{\Psi }\Psi =m\overline{\Psi }K_\mu \gamma ^\mu \Psi
\end{equation}
These identities have close relation with the modified normed condition (14).
}

Next we use the Dirac idea of non-integrable phases$^{[8]}$.

\theorem{The charged massive Dirac equation can be rewritten in the following
''uncharged massless'' forms 
\begin{equation}
\begin{array}{ll}
& i\gamma ^\mu (\partial _\mu -ieA_\mu )R-mL \\ 
= & i\gamma ^\mu (\partial _\mu -ieA_\mu +imK_\mu )R=[i\gamma ^\mu \partial
_\mu R_0]\Theta ^{-1}=0
\end{array}
\end{equation}
\begin{equation}
\begin{array}{ll}
& i\gamma ^\mu (\partial _\mu -ieA_\mu )L-mR \\ 
= & i\gamma ^\mu (\partial _\mu -ieA_\mu +imK_\mu )L=[i\gamma ^\mu \partial
_\mu L_0]\Theta ^{-1}=0
\end{array}
\end{equation}
where $R_0\stackrel{d}{=}R\cdot e^{-i\int (eA_\mu -mK_\mu )dx^\mu }=R\cdot
\Theta $.
}

It is equivalent to 
\begin{equation}
\begin{array}{ll}
& i\gamma ^\mu (\partial _\mu -ieA_\mu )\Psi -m\Psi \\ 
= & i\gamma ^\mu [\partial _\mu -i(eA_\mu -mK_\mu )]\Psi =[i\gamma ^\mu
\partial _\mu \Psi _0]\Theta ^{-1}=0
\end{array}
\end{equation}
here $\Psi _0\stackrel{d}{=}\Psi \cdot e^{-i\int (eA_\mu -mK_\mu )dx^\mu }$
satisfies the massless Dirac equation.

One knows that a charged particle must be massive. Thus in our formalism the
mass $m$ was introduced in the same way as the charge $e$! It is
mathematically beautiful and physically natural.

It is important to notice that in general $K_\mu =[\mbox{Re}(K_\mu )+
i\mbox{Im}(K_\mu )]$ is complex : 
\begin{equation}
\begin{array}{lll}
i\mbox{Im}(K_\mu ) & =\frac{(\overline{R}\gamma _\mu R-\overline{L}\gamma
_\mu L)(\overline{L}R-\overline{R}L)}{4(\overline{R}L)(\overline{L}R)} & =%
\frac{\pi _\mu ^5(\overline{\Psi }\gamma ^5\Psi )}{(\pi _\nu \pi ^\nu )} \\ 
\mbox{Re}(K_\mu ) & =\frac{(\overline{R}\gamma _\mu R+\overline{L}\gamma _\mu L)(%
\overline{R}L+\overline{L}R)}{4(\overline{R}L)(\overline{L}R)} & =\frac{\pi
_\mu (\overline{\Psi }\Psi )}{(\pi _\nu \pi ^\nu )}
\end{array}
\end{equation}
here 
\begin{equation}
\begin{array}{l}
\pi ^\mu =\overline{\Psi }\gamma ^\mu \Psi =G_\lambda ^{*}c^{\lambda \mu \nu
}G_\nu \\ 
\pi _5^\mu =\overline{\Psi }\gamma ^\mu \gamma ^5\Psi =-G_\lambda ^{*}\check{%
c}^{\lambda \mu \nu }G_\nu
\end{array}
\end{equation}
It means that $\mbox{Re}(K^\mu )$ is associated with unit vector $k^\mu $ and $%
\mbox{Im}(K^\mu )$ is associated with unit vector $j^\mu $.

Only the real part $\mbox{Re}(K_\mu )$ is associated with the phase, and the
imaginary part $\mbox{Im}(K_\mu )$ is associated with the ``scale factor'' $\sigma
=e^{-m\int [\mbox{Im}(K_\mu )]dx^\mu }$ of a spinor. (A similar scale change can be
found in the Weyl's earlier work.$^{[9]}$). Furthermore 
\begin{equation}
\lbrack \mbox{Re}(K_\mu )]\cdot [\mbox{Im}(K^\mu )]=k^\mu \eta _{\mu \nu }j^\nu =0
\end{equation}
\begin{equation}
\begin{array}{l}
\overline{\Psi }[\mbox{Im}(K_\mu )]\gamma ^\mu \Psi =0 \\ 
\overline{\Psi }[\mbox{Re}(K_\mu )]\gamma ^\mu \Psi =\overline{\Psi }K_\mu \gamma
^\mu \Psi =\overline{\Psi }\Psi
\end{array}
\end{equation}
One can realize that the trilinear form is independent of $\mbox{Im}(K_\mu )$.
Thus, if 
\begin{equation}
\Psi _0\stackrel{d}{=}\Psi \cdot e^{-i\int (eA_\mu -mK_\mu )dx^\mu }=\Psi
\Theta
\end{equation}
\begin{equation}
\begin{array}{lll}
\overline{\Psi }\stackrel{d}{=}\Psi ^{*T}\gamma ^0 & ; & \overline{\overline{%
\Psi _0}}\stackrel{d}{=}\overline{\Psi }\cdot \Theta ^{-1}
\end{array}
\end{equation}
then the Lagrangian 
\begin{equation}
\begin{array}{ll}
& \overline{\Psi }i\gamma ^\mu (\partial _\mu \Psi -ieA_\mu )\Psi -m%
\overline{\Psi }\Psi \\ 
= & \overline{\Psi }i\gamma ^\mu [\partial _\mu \Psi -ieA_\mu +im
\mbox{Re}(K_\mu )]\Psi \\ 
= & \overline{\overline{\Psi _0}}i\gamma ^\mu \partial _\mu \Psi _0
\end{array}
\end{equation}
is independent of the ``scale factor''
$\sigma =e^{-m\int [\mbox{Im}(K_\mu )]dx^\mu}$.

\section{Non-integrable exponential factors}

A physical spinor field $\Psi $ which satisfies the massive charged Dirac
equation (i.e. all physical solutions of massive Dirac equation) can be
expressed in the following form

\begin{equation}
\begin{array}{ll}
\Psi & \stackrel{d}{=}\Psi _0\cdot e^{i\int (eA_\mu -mK_\mu )dx^\mu } \\ 
& =R_0\cdot e^{i\int (eA_\mu -mK_\mu )dx^\mu }+L_0\cdot e^{i\int (eA_\mu
-mK_\mu )dx^\mu }
\end{array}
\label{59}
\end{equation}
Here $\Psi _0$ satisfies the {\it massless} Dirac equation and 
\begin{equation}
K_\mu =\frac{\overline{R}\gamma _\mu R}{2\overline{R}L}+\frac{\overline{L}%
\gamma _\mu L}{2\overline{L}R}=\frac{\overline{R_0}\gamma _\mu R_0}{2%
\overline{R_0}L_0}+\frac{\overline{L_0}\gamma _\mu L_0}{2\overline{L_0}R_0}
\label{60}
\end{equation}
The plane-wave solution of an electron is the most important solution in
the quantum field theory. In this special case, the real time-like
vector $mK_\mu$ is nothing but the energy-momentum of a massive
Dirac particle.

The connection between non-integrability of phase and electromagnetic
potential $A_\mu $ given here is not new, which is essentially just Weyl's
(1929) principle of gauge invariance$^{[10]}$ in its modern form. C.N.Yang
has reformulated the concept of a gauge field in an integral formalism and
extended Weyl's idea to the more general non-Abelian case$^{[11]}$. The
non-integrable phase for the wave functions was also discussed by Dirac in
1931$^{[8]}$, where the problem of monopole was studied. He emphasizes that
``non-integrable phases are perfectly compatible with all the general
principles of quantum mechanics and do not in any way restrict their
physical interpretation.'' Dirac conjectured that : ''{\it The change in
phase of a wave function round any closed curve may be different for
different wave functions by arbitrary multiplies of }$2\pi ${\it ''.} There
is the famous Dirac relation: $eq/(4\pi )=(n/2)$ . This means that if the
quantization of electric charge (the universal unit $e$ ) is accepted, then
the above relation is the law of {\it quantization} of the magnetic pole
strength.

Notice, in our case $K^\mu =[\mbox{Re}(K^\mu )+i\mbox{Im}(K^\mu )]$ is
complex. Thus only $i\int [eA_\mu -m(\mbox{Re}(K_\mu ))]dx^\mu $ is
associated with the phase change, and the imaginary part
$\int \mbox{Im}(K_\mu )dx^\mu $ is associated with the scale change.

Because of the single-valued nature of a quantum mechanical wave function,
we naturally conjecture that:

1. The{\it \ \underline{phase} change }of a wave function round any closed
curve must be close to $2n\pi $ where $n$ is some integer, positive or
negative. This integer will be a characteristic of possible singularity in $%
m \mbox{Re}(K_\mu )$.

2. The {\it \underline{scale} change }of a wave function round any closed
curve must be close to {\it zero}. As being mentioned in the previous
section, the Lagrangian (67) is independent of this scale change. This
Lagrangian in the massless form is conformally invariant$^{[12]}$. Thus
this scale-factor can be gauged away by conformal rescaling.

This is a new (very strong) assumption, and cannot be proved, not derived.
It is a conjecture of the overall consistency among all the solutions to the
same equation. The existence of magnetics monopole is an open question yet,
thus in our case it means that the change in $\oint mK_\mu dx^\mu $ round
any closed curve, with the possibility of there being singularity in $%
m \mbox{Re}(K_\mu )$ , will lead to the law of {\it quantization} of
physical constants, including {\bf mass}.

\end{document}